**Colossal negative magnetoresistance in dilute fluorinated graphene**

X. Hong[1], S.-H. Cheng[1], C. Herding[1,3] and J. Zhu[1,2*]

1. Department of Physics, The Pennsylvania State University, University Park, Pennsylvania, 16802, USA
2. The Materials Research Institute, The Pennsylvania State University, University Park, Pennsylvania, 16802, USA
3. Physikalisches Institut, Universität Münster, Wilhelm-Klemm-Str. 10, 48149 Münster, Germany

*Email: jzhu@phys.psu.edu

**Abstract**

Adatoms offer an effective route to modify and engineer the properties of graphene. In this work, we create dilute fluorinated graphene using a clean, controlled and reversible approach. At low carrier densities, the system is strongly localized and exhibits an unexpected, colossal negative magnetoresistance. The zero-field resistance is reduced by a factor of 40 at the highest field of 9 T and shows no sign of saturation. Unusual "staircase" field dependence is observed below 5 K. The magnetoresistance is highly anisotropic. We discuss possible origins, considering quantum interference effects and adatom-induced magnetism in graphene.

Defects and adsorbates have proven powerful in altering the electronic properties of graphene through doping, scattering and band gap induction [1-8]. In particular, point defects such as vacancies and adatoms can perturb the electronic states of graphene strongly, leading to midgap states [1] and drastic change of transport properties [4-8]. With the ability to control the form and density of defects and the assistance of microscopic imaging tools, defect-engineered graphene can be used as model systems to study the complex role of disorder in two dimensions. Moreover, it has been predicted that point defects may also introduce local magnetic moments and magnetic interactions unavailable in pristine graphene. Calculations show that the peculiar nature of Dirac fermions gives rise to ferromagnetic (antiferromagnetic) interactions among moments occupying the same (opposite) graphene sublattice and possible competing magnetic orders [9, 10]. Evidence of magnetism in single-layer graphene remains elusive to date. A controllable defect coverage together with *in situ* wide tunability of the electronic states makes defective graphene an ideal venue to explore the above novel physical phenomena.

In this Letter, we create clean, dilute fluorinated graphene (DFG) and report the observation of unexpected, colossal, negative magnetoresistance (MR). As the electron density is reduced to below the density of fluorine adatoms, the system undergoes a transition from weak to strong localization. In the strongly localization regime, a perpendicular magnetic field drastically reduces the DFG resistance by up to 40-fold at 9 T, while an-plane magnetic field causes only a very small positive MR. We compare our data to existing orbital mechanisms and discuss the possibility of magnetism in our samples.

Fluorination controls the properties of carbon materials effectively [11]. Fluorine adatoms are chemically simple and stable in ambient conditions. In this study, we work in the extreme dilute limit such that the properties of graphene dominate. Fluorine adatoms are covalently attached to graphene using $CF_4$ plasma [12]. This fluorination process is nearly completely reversible. We monitor the concentrations of fluorine adatoms and the unintentionally generated vacancies using Raman spectroscopy (Fig. 1a). With a carefully optimized recipe, the number of vacancies is minimized, as shown by the minimal D band in Fig. 1a taken on defluorinated DFG samples. The methods of fluorination and defluorination are described in detail in Ref. [13]. Defluorinated DFG samples exhibit mobility of ~2000 cm$^2$/Vs and Shubnikov-de Haas oscillations characteristic of high-quality single-layer graphene (Fig. 3d), confirming that F-adatoms are the dominant form of defects.

We employ scanning tunneling microscopy (STM) to measure the concentration and distribution of F-adatoms on graphene. The technical details are given in Ref. [13]. Representative images are shown in Fig. 1b. Isolated, three-fold symmetric, ($\sqrt{3}\times\sqrt{3}$)R30° superstructures are identified, indicating covalently bonded fluorine adtoms [14]. They are semi-ionic in nature at this dilute concentration [12]. Analyzing images covering a total area of 0.04 μm$^2$ and 900 defects, we estimate the average fluorine-fluorine spacing to be 7 nm, which corresponds to a concentration of $n_F$ = 2x10$^{12}$/cm$^2$ and a dilute F/C ratio of 1 to 2000, or 0.05%.

DFG samples are fabricated into field effect transistors in Hall bar geometries using standard e-beam lithography (Fig. 2a inset) [15]. Transport measurements are done in a pumped $^4$He cryostat equipped with a rotator and a 9 T magnet. We employ standard lock-in techniques and carefully select the excitation current of 0.5-50 nA to avoid Joule heating. Representative transport data taken on three fluorinated samples (denoted as A, B and C) and one defluorinated sample are shown here.

Figure 2a plots sheet resistsance $R_s(V_g)$ on sample A at $B$ = 0 T and selected temperatures. Surprisingly, even with such a dilute F concentration, $R_s(V_g)$ is remarkably different from that of pristine graphene, which shows weak and mostly metallic $T$–dependence. Here, the resistance at the charge neutrality point increases by three orders of magnitude from 25 kΩ at 200 K to 2.5 MΩ at 5 K, displaying a strong insulating behavior. This insulating $T$–dependence persists to higher carrier densities but becomes progressively weaker with increasing $n$. From 200 K to 5 K, $R_s$ at hole density $n$ = 4.2x10$^{12}$/cm$^2$ only changes from 5 kΩ to 7 kΩ. Electron-side and hole-side show approximately symmetric gate dependence. Similar $R_s(V_g)$ are also observed in hydrogenated and ozone-damaged graphene [4, 5, 16].

$R_s(T)$ at different carrier densities show three distinct behavior, consistent with strong localization (SL), weak localization (WL) and a transition in between. Figure 2b show representative curves at different densities in sample B. In the SL regime, $R_s(T)$ follows the $T$–dependence of Mott's two-dimensional (2D) variable range hopping (VRH) [17]:

$$R_s \propto \exp(T_0/T)^{1/3}; \text{ and } T_0 = \frac{13.8}{k_B N(E_F) \xi^2} \propto E_b^{1.5} = (E_c - E_F)^{1.5} \qquad (1).$$

Here $E_c$ is the mobility edge, $E_F$ is the Fermi level and $E_b$ the binding energy. $N(E_F)$ is the density of localized states at $E_F$. $T_0$ is the characteristic temperature and $\xi$ the localization length. $T_0$ = 280 K at the charge neutrality point and decreases with increasing $n$ (Fig. 2b inset). At $n$ = 1.4x10$^{12}$/cm$^2$, $T_0$ reaches 24 K and only the low temperature $R_s$ ($T$ < 100 K) can still be described by the VRH model (green triangles in Fig. 2b). As $n$ further increases, $R_s(T)$ deviates from the $\exp(T_0/T)^{1/3}$ expression at progressively lower temperature. For $n$ > 2.5x10$^{12}$/cm$^2$, $R_s(T)$ can no longer be described by the VRH model but instead exhibits a logarithmic $T$–dependence [18], consistent with weak localization [19].

Such density-driven SL to WL transition is seen in all three DFG samples. The $n$–dependence of $T_0$ is given in the inset of Fig. 2b. The apparent crossover from $\exp(T_0/T)^{1/3}$ to $\ln T$ dependence occurs in the vicinity of $R_c \sim 12$ k$\Omega/\square$ ($\sim h/2e^2$) (grey band in Fig. 2b). The transition density regime where $R_s(T)$ crosses $R_c$ is roughly 1.5-2.5x10$^{12}$/cm$^2$, where $T_0$ approaches zero. The transition densities are comparable to the average fluorine density $n_F = 2\times10^{12}$/cm$^2$ in these samples and points to the important role of the F-adatoms in this transition. Calculations based on midgap states [1] can account for the $R_s(V_g)$ behavior at high temperature [18]. However, the temperature and density dependence of $R_s$ in our samples are far more complex than suggested by existing calculations for this dilute fluorine coverage [7, 8].

Most strikingly and surprisingly, in a magnetic field perpendicular to the graphene plane, DFG samples display colossal negative magnetoresistance (MR) in the SL regime. Figure 3a plots the normalized MR, $R_s(B)/R_s(0)$, in sample A for three densities in the SL, transition and WL regimes, respectively. While all show negative MR, both the $B$-dependence and the magnitude of the MR are drastically different. In the WL regime, the negative MR is significantly smaller, decreases sharply at low field and quickly saturates. Its $B$-dependence is well described by the theory of weak localization [13, 19]. $R_s(B)$ in the SL regime exhibits a parabolic $B$-dependence at low field, followed by large negative MR with no sign of saturation at 8.9 T. We observe up to 40-fold reduction in resistance ($R_s(9T)/R_s(0) = 0.025$) in this regime and have verified that weak localization cannot describe the data [13]. A complex scenario arises for densities in the transition regime, where the MR appears to carry characteristics of both WL and SL. Next, we focus on the large MR in the SL regime [16].

Figure 3b plots $R_s(B)/R_s(0)$ at selected temperatures and $n = 8\times10^{11}$/cm$^2$ in sample A. Above 5 K, $R_s(B)/R_s(0)$ decreases smoothly with increasing $B$, with increasing MR at lower temperature. Intriguingly, below 5 K, a "staircase" $B$–dependence starts to develop, where $R_s(B)/R_s(0)$ alternates between saturation and sharp decrease. This staircase behavior becomes more pronounced at lower temperature and appears over a wide density range in the SL and transition density regimes in multiple samples (Fig. 3b inset). Such behavior has not been reported in strongly localized systems before.

To further illustrate the role of the magnetic field, we plot in Fig. 4a $R_s(T)$ at fixed $B$ values for the data shown in Fig. 3b. $R_s(T)$ is well described by 2D VRH conduction at all $B$ fields, but the application of a magnetic field substantially reduces the degree of localization, with the extracted $T_0(B)$ decreasing from 330 K at 0 T to 26 K at 8.9 T (Fig. 4b). In Fig. 4b, we also plot the $B$-dependence of the localization length $\xi(B)$, which is calculated using Eq. 1 and assuming a $B$-independent $N(E_F)$ approximated by the density of states of graphene at this density. The localization length $\xi(B)$ increases by a factor of 3.6 from 56 nm at 0 T to 200 nm at 8.9 T, and shows no signs of saturation.

Strongly localized systems exhibit a rich variety of MR, both positive and negative, the majority of which is much smaller than what is observe here. Negative MR of this magnitude was seen in disordered 2D electron gas leading to the re-entrant quantum Hall state [20, 21]; however, the $B$-dependence of the MR there looks very different and the phenomenon is attributed to a structural change of the density of states in the magnetic field [20, 21]. Zeeman effect can also lead to a large negative MR in systems near the metal-insulator transition [22], but the effect is isotropic. Figure 3c shows the magnetoresistance of sample B in an in-plane field. A small (< 4%), positive MR is seen at 9 T, in stark contrast to the large negative MR seen in perpendicular fields. The large anisotropy rules out spin-related mechanisms and indicates the orbital origin of the negative MR. In strongly localized systems, two quantum interference effects may produce negative MR. The first considers the interference of forward hopping paths enclosing an area $A_{\text{NSS}} = (r_m)^{3/2} \xi^{1/2}$, where $\xi$ is the localization length given in Eq. (1) and $r_m = \xi (T_0/T)^{1/3}$ is the hopping distance and $r_m \geq \xi$ [23]. This effect can only produce a negative MR of less than 10% and should saturate near $B^* A_{\text{NSS}} = \phi_0 = h/e$. In Fig. 4b, an upper bound of $B^*$ using $r_m = \xi$ yields 0.21 T. This mechanism clearly cannot explain our data. In the second interference effect [24], the magnetic field breaks the time reversal symmetry between forward and backward hopping paths, in analogy to weak localization in metallic systems, and causes an increase of $\xi$ and negative MR. In quasi-1D systems, $\xi(B)/\xi(0)$ was shown to saturate at a universal value of 2 at $B > B^*$. It was later pointed out in Refs. [25, 26] that in higher dimensions $\xi(B)/\xi(0)$ is no longer a universal number, but the theory in this regime remains to be fully developed. Experimental evidence of this effect is scant [24, 27], with the doubling of $\xi$ shown in Ref. [24] in three dimensional GaAs and the largest enhancement of conductance being ten-fold in $In_2O_{3-x}$ films [27]. Both systems [24, 27] show a saturation of MR around $B^* = \phi_0/\xi^2$. This mechanism can potentially explain the large negative MR seen in our samples since it was argued that $\xi(B)$ can be enhanced by an arbitrary factor in two and three dimensions [25]. However, several notable differences exist between our data and past experiments and theory. Firstly, in our samples, $(\xi(B^*) - \xi(0))/\xi(0)$ is ~ 0.5%, compared to the order of unity predicted by theory [24-26] and observed in Refs. [24, 27]. Secondly, $\xi(B)$ shows no sign of saturation at the highest field 8.9T where the magnetic length $l_B = (h/eB)^{1/2} = 8.5$ nm becomes much smaller than $\xi(0) = 56$ nm; on the contrary, $d^2\xi(B)/dB^2 > 0$. Lastly and very importantly, the "staircase" MR seen in our samples at low temperature seems to suggest the existence of discrete energy levels, which can be probed by the magnetic field. This behavior has not been reported in any strongly localized systems before and appears to be difficult to reconcile with existing theories, which lack such features. Interestingly, $l_B$ = 8.5 nm at 9 T becomes comparable to the mean free path ($l_e$ = 5 nm at 200 K) and brings in the tantalizing possibility of the Hofstadter regime [24]. If the large negative MR in our samples indeed originates from quantum interference effects, it is quite extraordinary that nearly 100% of the resistance of our samples comes from interference. The unusual parameter regime ($l_e \leq l_B << \xi_0$) our data probe certainly presents several challenges and opportunities to further theoretical studies on this subject. In this regard, dilute fluorinated graphene offers a fresh venue to explore the rich phenomena of localization in 2D in the context of Dirac Fermions.

Motivated by the "staircase" MR and the strong resemblance of our data to those reported for colossal magnetoresistive manganites and dilute magnetic semiconductors on the insulating side close to the ferromagnetic/metal-insulator transition [28-30], we propose a second scenario based on the formation of magnetic polarons and the delocalization effect of the magnetic field [29]. It has been predicted theoretically that magnetic moments of size $\sim \mu_B$ can be induced in graphene via adatoms, substitutes or vacancies [9, 10]. At the present, there is no direct evidence of magnetism in single-layer graphene, nor is direct magnetometric measurement possible due to the limited size of most graphene samples.

In the magnetic polaron model, a magnetic polaron forms between a localized electron spin **s** and nearby local moments **S**. The exchange coupling $J$ **S·s**, being either ferromagnetic or antiferromagnetic, enhances the bare binding energy of localized electrons $E_{b0}$ to $E_b{}^* = E_{b0}+E_{ex}$ [29]. The alignment of polarons in an external magnetic field enhances their hopping probability, thereby reducing $E_b{}^*$ and causing the polarons to unbind. Although the dimensionality, band structure, and the origin of the magnetic moments vary considerably among manganites, semiconductors and graphene, the core ideas of the magnetic polaron picture is appealing, given the concurrent onset of strong localization and colossal negative MR in our DFG samples. It can also account for the observed negative MR as a consequence of unbinding polarons in a magnetic field.

Applying the above concepts to our samples, we obtain a lower bound of 7 meV for the enhanced binding energy $E_b{}^*$ at $B = 0$ [13]. At 9 T, $E_b{}^*$ is reduced by a factor of 5, using the $T_0(B)$ data shown in Fig. 4b and the scaling in Eq. (1). This indicates that exchange contribution dominates the polaronic binding energy.

In the polaron scenario, the absence of large MR in an in-plane magnetic field suggests a perpendicular orientation of the local moments. This observation is consistent with predictions of enhanced spin-orbit coupling of a few meV near adatoms, as a result of σ−π mixing [10, 31, 32]. If true, the large magnetic anisotropy may be explored in spintronics applications to control the relaxation of spins in different orientations.

Independent of the origin of the colossal negative MR, the magnitude of the phenomena provides the high sensitivity necessary for magnetic readers. In both scenarios, the temperature and field dependence of the MR are not well understood. The "staircase" $B$-dependence at low temperatures (Fig. 3b) is particularly perplexing. Interestingly, "staircase" $B$−dependence is also seen in magnetization simulations of anti-ferromagnetically coupled local moments occupying opposite graphene sublattices [10].

Despite the dilute concentration, F-adatoms within 2 nm of each other are observed (Fig. 1b inset). These unanswered questions call for more experimental and theoretical studies.

In conclusion, we demonstrate that fluorination provides a clean, reversible and tunable approach to engineer defects and control the properties of graphene. Dilute fluorinated graphene exhibits unexpected, anisotropic, colossal negative magnetoresistance and unusual "staircase" behavior at low temperature. This new material system offers fascinating opportunities to examine mesoscopic phenomena in the context of Dirac Fermions and explore potential magnetic interactions.

**Acknowledgements**

We are grateful for helpful discussions with B. Altshuler, A. Ayuela, A. H. Castro Neto, V. Crespi, M. Fuhrer, T. Rappoport, N. Samarth, P. Schiffer, J. Sofo, and B. Uchoa, and technical assistance from H. Gutierrez, J. Li, R. Misra, P. Joshi, M. Tian, and B. Wang. We thank N. Samarth and P. Schiffer for providing access to their Vecco multi-mode system. Work at Penn State is supported by NSF Grants CAREER No. DMR-0748604, NIRT No. ECS-0609243 and MRSEC No. DMR-0820404. The authors acknowledge use of facilities at the PSU site of NSF NNIN.

**Figure Captions:**

Fig. 1 a) Comparison of Raman spectra on fluorinated (upper trace) and defluorinated (lower trace) graphene. Fluorination conditions are identical for both samples. The Raman bands are marked in the figure. b) A STM current image (12 nm × 12 nm) on fluorinated graphene. $V_{bias}$ = 20 mV. Color scale: 2 nA. The slowly varying background is due to the height undulation of the SiO$_2$ surface. The honeycomb lattice of unperturbed graphene is marked by white solid lines. The white dotted lines mark the signature of isolated fluorine adatoms [14]. The fluorine distribution is inhomogeneous. F-adatoms within 2 nm of each other are sometimes observed. Clusters are rare. Inset: three F-adatoms close to each other with overlapping superstructures. Color scale: 5 nA.

Fig. 2 a) Semi-log plot of $R_s$ vs. backgate $V_g$ on sample A at $T$ = 5, 10, 20, 50, 100, 200 K (top to bottom). The backgate dopes $7 \times 10^{10}$/cm$^2$/$V_g$(V) and the charge neutrality point occurs at $V_g$ = 16 V. Inset: optical image of a DFG device with the graphene piece outlined. b) Semi-log plot of $R_s$ vs. $T^{-1/3}$ on sample B at the charge neutrality point, $n$ = 0.7, 1.4, and $2.5 \times 10^{12}$/cm$^2$ (top to bottom). Dashed lines are fittings to the VRH conduction, the slope of which yields $T_0$. The grey band corresponds to ~ $h/2e^2$. Inset: The $n$−dependent $T_0$ for samples A (black squares), B (red circles), and C (green triangles) in the strong localization regime.

Fig. 3 a) Normalized MR in a perpendicular magnetic field, $R_s(B)/R_s(0)$, on sample A at $T$ = 5 K and $n$ = 0.8 (black), 2.1 (red), and $4.2 \times 10^{12}$/cm$^2$ (blue). Each curve represents a distinct regime as labeled. b) Normalized MR at $n$ = $0.8 \times 10^{12}$/cm$^2$ and selected temperatures. Inset: staircase MR observed on another sample (B) at $n$ = 2.1 (navy, upper) and $1.4 \times 10^{12}$/cm$^2$ (orange, lower). $T$ = 2 K. c) Normalized MR on sample B at the charge neutrality point in perpendicular ($B_\perp$) and in-plane ($B_{//}$) magnetic field showing a large anisotropy. d) Comparison of magnetoresistance of DFG (black, scale on the left) and defluorinated DFG (red, scale on the right). For both samples, $n = 1.05\times10^{12}$/cm$^2$ and $T =$ 5 K. The defluorinated sample exhibits magnetoresistance oscillations with the filling factors marked in the figure.

Fig. 4 a) $R_s$ vs. $T^{-1/3}$ in a semi-log plot for data shown in Fig. 3b at selected fields. Solid lines are fittings to the VRH conduction, the slope of which yield $T_0$. b) $T_0$ (scale on the left) and corresponding $\xi$ (scale on the right) vs. $B$ from the data shown in a). The blue arrow marks $B^* = 0.21$ T.

# Figure 1

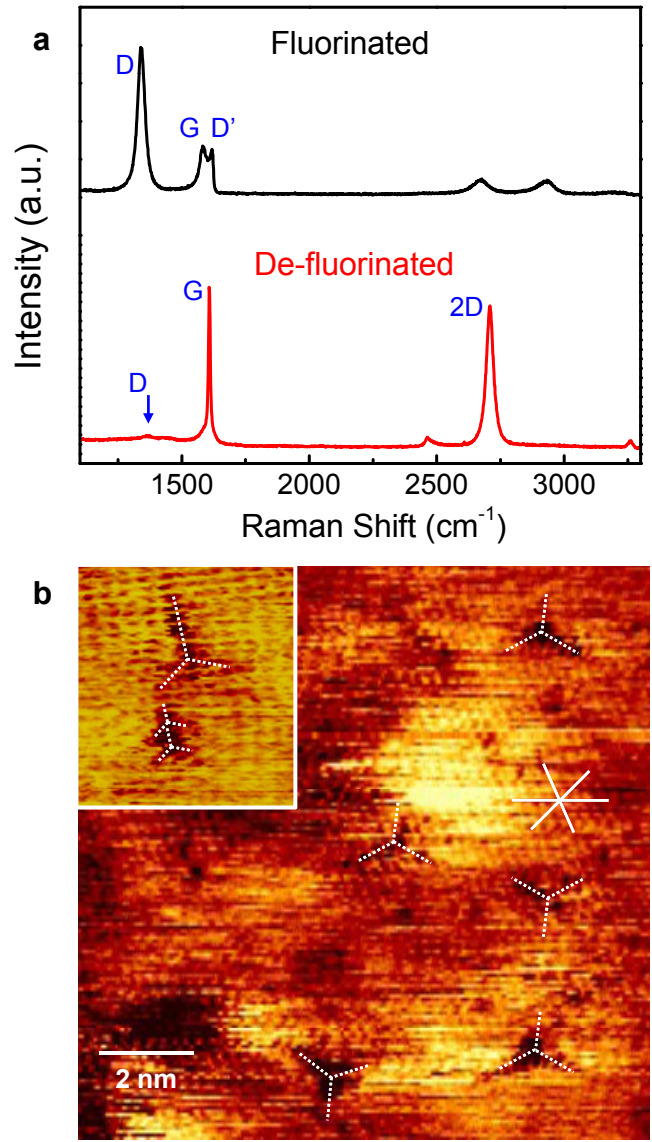

Figure 2

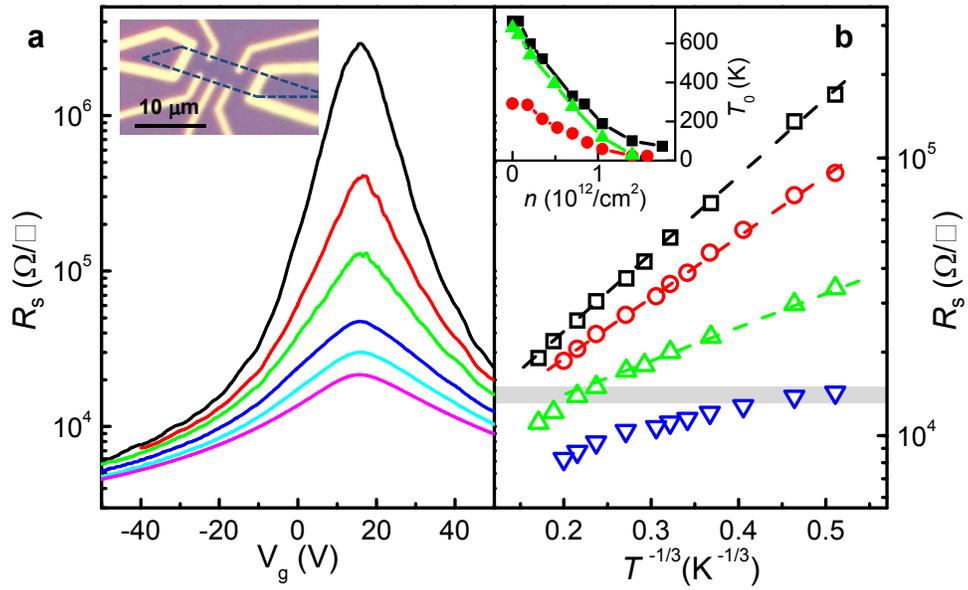

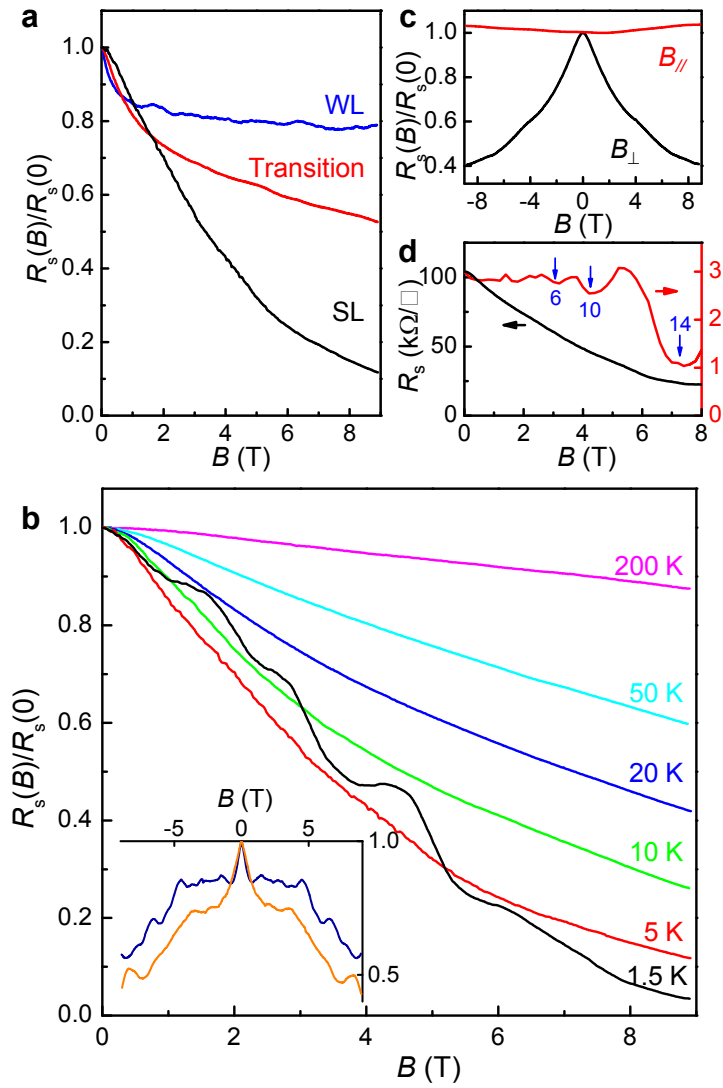

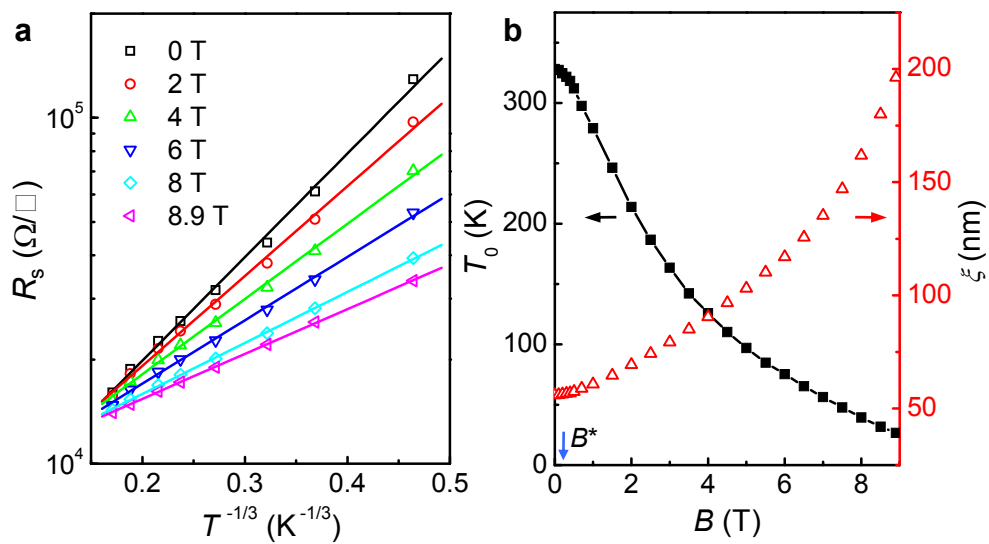

**Colossal negative magnetoresistance in dilute fluorinated graphene (Supplementary Information)**

X. Hong[1], S.-H. Cheng[1], C. Herding[1,3] and J. Zhu[1,2]

1. Department of Physics, The Pennsylvania State University, University Park, Pennsylvania, 16802, USA
2. The Materials Research Institute, The Pennsylvania State University, University Park, Pennsylvania, 16802, USA
3. Physikalisches Institut, Universität Münster, Wilhelm-Klemm-Str. 10, 48149 Münster, Germany

Correspondence to J. Zhu, Email: jzhu@phys.psu.edu

**Supplementary Information Content**

1. Sample Preparation and Characterization
2. Weak Localization Fittings to the Magnetoconductance at High and Low Densities
3. Estimate of Magnetic Polaron Binding Energy

1. **Sample Preparation and Characterization**

   Graphene sheets on $SiO_2$/Si substrates are fluorinated in a reactive ion etching system (PlasmaTherm 720) using $CF_4$ plasma at room temperature for 5 to 30 minutes. The $CF_4$ gas pressure is 100 mTorr and the power is set to 5 W. Fluorinated graphene is stable in ambient and through lithographic processes as indicated by Raman spectroscopy. To remove fluorine, we anneal fluorinated graphene sheets in a tube furnace in the flow of forming gas (90% Ar/10% $H_2$, 25 sccm) for 24 hours at 365°C.

   For STM studies, gold electrodes are evaporated on fluorinated graphene sheets via a shadow mask. STM measurements are carried out on a Veeco Multimode system equipped with a home-made acoustic enclosure to improve image quality. We use commercial Precision-cut Platinum/Iridium STM probes from Veeco. Current images are taken with a tip bias of $V_{bias} = 20$ mV. The tunneling current is typically a few nA.

   Raman data are taken with a micro-Raman spectrometer (inVia, Renishaw, Inc.) using 514 nm excitation wavelength.

2. **Weak Localization Fittings to the Magnetoconductance at High and Low Densities**

   Figure S1 shows the conductance of Sample A as a function of magnetic field at two representative carrier densities. At a high carrier density of $n = 4.2 \times 10^{12}/cm^2$, the magnetoconductance can be well described by the theory of weak localization [1], as the fitting in Fig. S1a shows. In contrast, the magnetoconductance at a lower density $n = 1.05 \times 10^{12}/cm^2$, where the sample exhibits the *T*-dependence of strong localization at zero magnetic field, deviates significantly from the weak localization behavior, as shown by the attempted fittings in Fig. S1b.

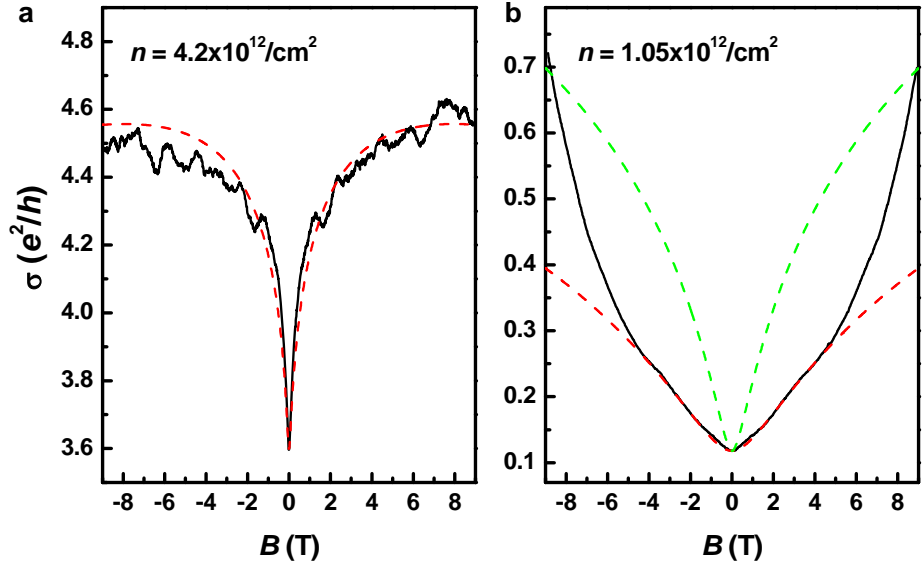

Fig. S1 Magnetoconductance of Sample A at 5 K. a) $n = 4.2 \times 10^{12}/cm^2$ and $\sigma(0\,T) > 2e^2/h$. The red dashed line is a fitting curve using the weak localization theory given in Ref. [1] with $\tau_\phi = 3.2$ ps. b) $n = 1.05 \times 10^{12}/cm^2$ and $\sigma(0T) \ll 2e^2/h$. The red and green fitting curves correspond to $\tau_\phi = 1.4$ ps and $\tau_\phi = 6.5$ ps, respectively. It is clear that weak localization theory does not capture the $B$-dependence at this density.

### 3. Estimate of Magnetic Polaron Binding Energy

We obtain a lower bound of the exchange enhanced magnetic polaron binding energy $E_b^*$ at $B = 0$ by noting that in the VRH regime, $E_b^* = E_c - E_F$ is larger than the varying activation energy of the bound polaron $E_a(T)$ given by [2]:

$$E_a(T) = d(\ln\rho)/d(k_B T)^{-1} = 1/3\, k_B (T_0 T^2)^{1/3} \tag{S1}$$

In Fig. 4a, $E_a(T)$ at $B = 0$ ranges from 80 K to 9 K. This suggests an $E_b^*$ ($B = 0$) > 7 meV, consistent with the formation of polarons in the temperature range studied. At 9 T, $E_b^*$ is reduced by a factor of 5, indicating that exchange contribution dominates the polaronic binding energy.